\documentclass{egpubl}
\usepackage{amsmath}
\usepackage[T1]{fontenc}
\usepackage{dfadobe}  
\usepackage{subcaption}
\usepackage{xcolor}
\usepackage{multirow}
\usepackage{tabularx}
\usepackage{array}
\newcolumntype{Y}{>{\centering\arraybackslash}X}
\definecolor{cemgreen}{rgb}{0.05, 0.6, 0.05}

\usepackage{cite}  
\BibtexOrBiblatex
\electronicVersion
\PrintedOrElectronic
\ifpdf \usepackage[pdftex]{graphicx} \pdfcompresslevel=9
\else \usepackage[dvips]{graphicx} \fi

\usepackage{egweblnk} 

\title[ ]%
      {HRsR: Hierarchical Rotation System Reconstruction}

\author[R. Cui, C. Akarsubaşı, E. T. Gæde, E. Rotenberg, L. Kobbelt, and J. A. Bærentzen]
{\parbox{\textwidth}{\centering Ruiqi Cui$^{1}$\orcid{0000-0003-1993-5036} Cem Akarsubaşı$^{1}$ Emil Toftegaard Gæde$^{1}$\orcid{0009-0001-9462-6359} Eva Rotenberg$^{2}$\orcid{0000-0001-5853-7909} Leif Kobbelt$^{3}$\orcid{0000-0002-7880-9470} J. Andreas Bærentzen$^{1}$\orcid{0000-0003-2583-0660}
}
        \\
{\parbox{\textwidth}{\centering $^1$Department of Applied Mathematics and Computer Science, Technical University of Denmark, Denmark\\
        $^2$ IT University of Copenhagen, Denmark\\
         $^3$Visual Computing Institute, RWTH Aachen University, Germany
       }
}
{\parbox{\textwidth}{\centering 
       }
}
}

\begin{document}


\maketitle
\begin{abstract}
   Surface reconstruction from point clouds remains challenging when both geometric fidelity and topology control are required. Rotation System Reconstruction (RsR) reconstructs triangle meshes from point clouds while explicitly controlling topology through the Euler characteristic, but its sequential edge insertion limits scalability. We present Hierarchical Rotation System Reconstruction (HRsR), which accelerates RsR through a hierarchical pipeline of edge collapses and vertex splits. HRsR first simplifies the input using a $k$-nearest neighbor graph, performs reconstruction on the reduced structure, and then restores geometric detail while preserving topology. To maintain geometric consistency, we incorporate intersection handling and quality-driven vertex split selection. Experiments demonstrate up to a $6\times$ speedup and more than $8\times$ reduction in memory usage over RsR, while achieving comparable reconstruction results.


\end{abstract}  

\section{Introduction}


Reconstructing a surface from a set of sampled 3D points is a challenging and ill-posed problem that entails inferring both geometry and the underlying topology from unorganized data that are usually acquired through optical acquisition technologies such as LiDAR, structured light scanning, or multi-view stereo. These methods produce point clouds which often suffer from incomplete coverage, noise, non-uniform sampling, and other artifacts which complicate the reconstruction process. 

Perhaps because of its wide-ranging downstream applications (including rendering, simulation, reverse engineering, shape analysis, etc.) a large number of methods for point cloud reconstruction have been proposed over the course of the past four decades. Most of these methods, including ours, produce triangle meshes as the output, and they can be divided into two categories: combinatorial methods which form triangles from the input points and methods which produce triangles that only approximate the input points. Methods from the latter category are often volumetric methods which work by constructing an implicit representation. From this implicit representation, a triangle mesh is extracted using iso-surface contouring. This approach has often been preferred since it tends to work well even if the input point cloud is noisy and incomplete. However, the noise resilience is largely due to the fact that volumetric methods introduce smoothing which is not necessarily desirable. Moreover, volumetric methods generally do not provide control over the topology of the resulting surface.


Recently, Cui et al. \shortcite{cui2024} proposed rotation system reconstruction (RsR), a combinatorial method which includes almost all points in the reconstructed mesh. The key idea behind RsR is that a tree (in the graph theoretical sense) is always a planar graph, which can be embedded in a genus 0 surface. Hence, forming a spanning tree of the input points yields a genus 0 reconstruction which only needs to be triangulated in order to produce the output. As long as the inserted edges split existing faces, the surface remains genus 0, and by adding handle edges the genus can be raised. RsR affords topology control by allowing the user to choose whether handle edges may be inserted.
\subsection{Contributions}
While RsR is effective, splitting faces is initially costly for large inputs since it starts from a mesh with a single face that includes all vertices. To mitigate this issue, we propose a hierarchical scheme which makes RsR efficient by collapsing a large fraction of the edges in an initial k-nearest neighbor graph before running RsR to produce a triangle mesh from the simplified graph. Finally, we add the details back onto the mesh by splitting vertices in the inverse order of the edge collapses. This hierarchical scheme has several important benefits compared to RsR.

\begin{trivlist}
\item{\textbf{Efficiency}}. On large point clouds, RsR is expensive in the early stages. By reducing the size of the input, we obtain a speedup of up to 6$\times$. Likewise, space efficiency is improved up to more than 8 $\times$.
\item{\textbf{Filtering}}. The edge collapse step ensures an even distribution of input points for RsR, obviating the need for removing almost coincident points. Moreover, during vertex splitting we can put a threshold on the lengths of edges introduced by splitting in order to avoid reintroducing those almost coincident points in the output or simply to produce an output at slightly coarser level of detail.
\end{trivlist}

\subsection{Related Work}
Our work draws inspiration from progressive meshes \cite{Hoppe1996}, but in Hoppe's work, edge collapse and vertex split are inverse operations and both applied to triangle meshes in order to build a hierarchical representation. In contrast, our method applies edge collapses to a graph to obtain an efficient starting point for RsR reconstruction, and vertex splits are used to reintroduce detail. Consequently, in our work, the vertex split operation is not the inverse of the edge collapse operation; it is merely needed to ensure that the manifoldness of the mesh is preserved.

As discussed above, existing surface reconstruction methods can be divided into two main categories: \textit{combinatorial} and \textit{volumetric}. Combinatorial methods directly interpolate the points, while volumetric methods approximate the surface by estimating a scalar function over 3D space, which encodes either an intensity value or a distance to the surface based on the point set.

\subsubsection{Volumetric reconstruction}
Volumetric methods have been extensively studied over the last few decades. A prominent example is Poisson reconstruction together with its variants \cite{kazhdan2006poisson, kazhdan2013screened, kazhdan2020poisson}. By solving the Poisson equation, Poisson reconstruction methods compute an implicit function whose gradient field best matches the input normals, from which the target surface is extracted as an isosurface. Other volumetric approaches address specific challenges in the input data. For instance, VIPSS and its variants \cite{xia2025variational, huang2019variational} deal with sparse, non-uniformly distributed, and unoriented inputs by solving a constrained quadratic optimization problem. More recently, the rise of deep learning has led to neural implicit methods for surface reconstruction \cite{erler2020points2surf, boulch2022poco}, which learn continuous surface representations directly from point cloud data.

However, volumetric methods inherently smooth the reconstructed surface, which in certain cases is undesirable. This limitation motivates RsR to adopt a combinatorial approach.

\subsubsection{Combinatorial reconstruction} 
Prior to RsR, a variety of combinatorial reconstruction methods have been proposed. An important class consists of Delaunay-based (or Voronoi-based) approaches, including Crust\cite{amenta1998new}, Power Crust\cite{amenta2001power}, CoCone\cite{amenta2000simple}, Co3Ne\cite{boltcheva2017surface}, restricted Delaunay triangulation\cite{Wang22}, etc. Through successive developments, these methods have become increasingly effective and robust in recovering geometry under challenging conditions such as sparse sampling, thin structures, and noisy input. However, in general, they do not provide guarantees on topological correctness. 
Another well-known combinatorial approach is the ball-pivoting algorithm (BPA) proposed by Bernardini et al.\cite{bernardini1999ball}. Starting from a seed triangle, BPA incrementally adds triangles by pivoting a ball along the point set. Digne et al.\cite{Digne2011} further improved this method by introducing a scale-space framework to better handle noise. Nevertheless, BPA-based methods may still produce incomplete reconstructions when faced with challenging inputs.


\subsubsection{Topology-aware surface reconstruction}
%
%
Persistent homology \cite{edelsbrunner2002topological} has been used as a tool for topology-aware reconstruction. Brüel-Gabrielsson et al. \cite{bruel2020topology} proposed a basis-function-based reconstruction method that incorporates topological constraints derived from persistence diagrams (PDs) into the optimization process. Similarly, Dong et al. \cite{dong2022topology} formulate reconstruction using B-spline functions, where topology is guided by a target function defined through PDs. 
Though powerful, persistence-based methods require the definition of a complex of primitives ordered according to a filtration. We build on RsR \cite{cui2024} whose topology control is based on how many handle edges that are allowed.

To achieve more flexible and precise control, Sharf et al. \cite{sharf2007interactive} introduced a user-interactive approach that allows users to specify local topology prior to fine-scale reconstruction. In addition, Sharf et al. \cite{sharf2006competing} proposed a deformable model that evolves from a genus-0 surface, where topology is controlled by allowing or preventing the merging of evolving fronts, thereby enabling the creation of handles or preserving the existing topology.
%
\begin{figure*}[htb]
  \centering
  \includegraphics[width=\linewidth]{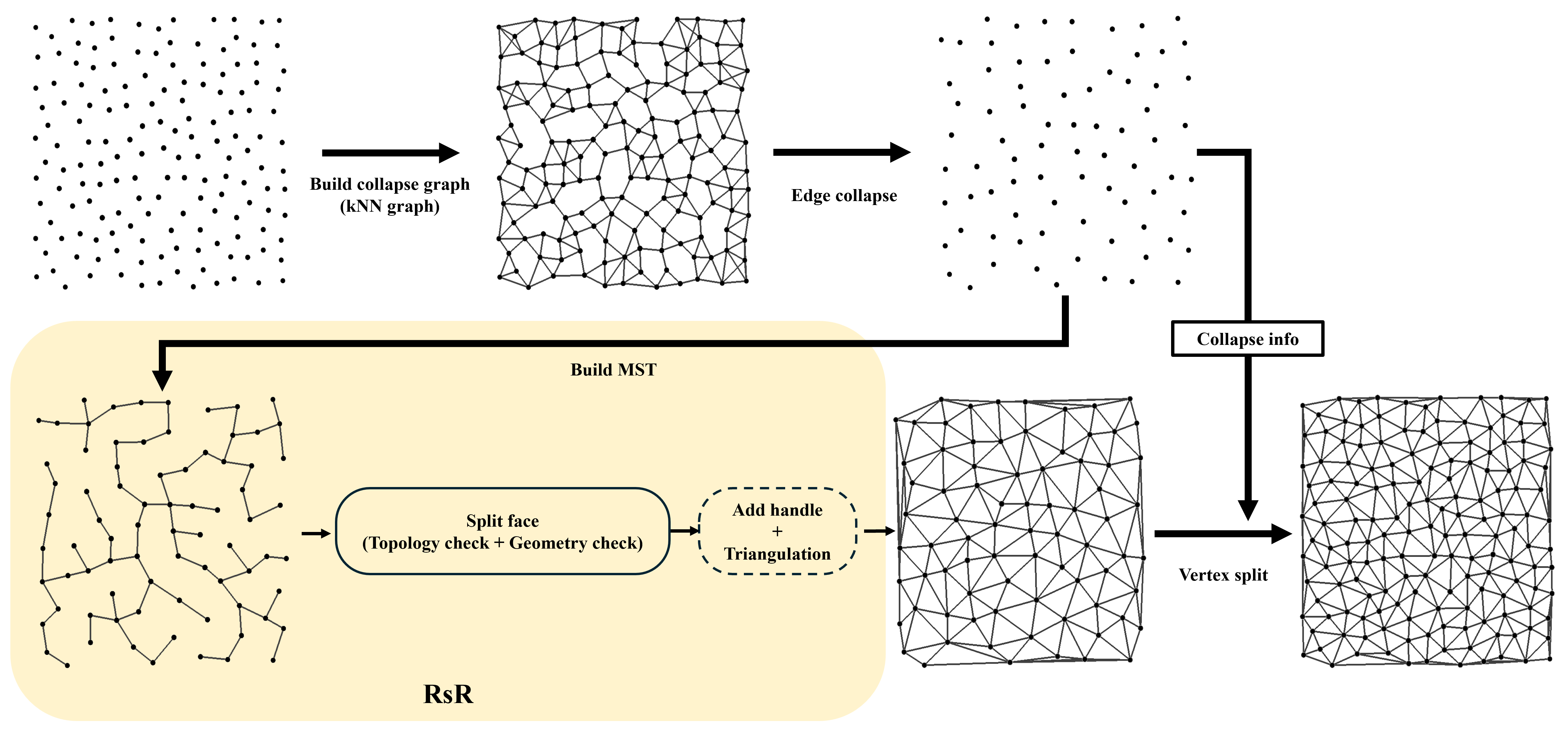}
  \caption{\label{fig:pipe}
           Pipeline of HRsR. HRsR starts with a simplification of the input point cloud by edge collapse. Subsequently, RsR is performed on the coarser point cloud. After reconstruction, we recover the original point density using the information we stored about the collapses.}
\end{figure*}


%
\section{Preliminaries}
Before getting into technical detail about HRsR, we briefly review RsR, the prior work we build upon, and define the new operators used in HRsR.
\subsection{Rotation system reconstruction}
RsR is a combinatorial surface reconstruction method that reconstructs triangle meshes from point clouds. The overall pipeline is illustrated in Figure \ref{fig:pipe}. Unlike the previously reviewed methods, RsR achieves topology control by explicitly tracking the Euler characteristic throughout the reconstruction process. 

The Euler characteristic, $\chi$, is given by 
\begin{equation}
    \chi = 2(1-g)
\end{equation}
where $g$ is the genus of the object. The Euler characteristic abides by the Euler-Poincaré formula,
\begin{equation}
\label{eq:euler}
    |V|-|E|+|F|= \chi = 2(1-g)
\end{equation}
where $|V|$ is the number of vertices, $|E|$ is the number of edges, $|F|$ is the number of faces. 

In edge-based mesh representations, $|V|$ and $|E|$ are always explicitly defined while $|F|$ is given by the number of closed edge cycles. To facilitate finding these cycles, RsR employs a \textit{rotation system} defined on top of the \textit{halfedge} representation. The rotation system specifies the cyclic ordering of outgoing halfedges for each vertex $v \in V$, which enables the determination of faces induced by the combinatorial structure. We refer readers to the original paper \cite{cui2024} for a more formal and detailed definition of halfedge representation and face construction under this representation.

Mesh editing operations which change the number of vertices, edges, or faces while maintaining \eqref{eq:euler}, are known as \textit{Euler operators}. If $g$ is unchanged by an operator, it is genus preserving, and otherwise it is genus changing. 
RsR uses two {Euler operators}. The genus preserving operation \textit{split face} inserts an edge that splits a single face into two. In this operation, both $|E|$ and $|F|$ increase by one while the genus doesn't change. Thus, Eq. \ref{eq:euler} remains true. The second Euler operator is \textit{add handle}. As the name suggests, the operator increases $g$ by one. This operation introduces a new edge while merging two faces into one, resulting in a net decrease of two in the left-hand side of Eq. \ref{eq:euler}, consistent with the change in genus.

Starting from a point cloud, RsR initially builds a $k$-nearest neighbor ($k$NN) graph and then a minimum spanning tree, which corresponds to a genus-0 mesh with a single face. RsR proceeds to incrementally reconstruct the surface by repeatedly applying face split to edge candidates in ascending order of length, forming a polygonal mesh where most faces have a low number of sides. Once the mesh has largely been reconstructed, RsR starts adding handles selectively, guided by the user input and a heuristic designed to distinguish spurious handles from valid ones. Finally, a triangulation step is performed to close the remaining minor holes and thereby complete the reconstruction. Holes which cannot be closed because the requisite edges are not in the input graph remain as holes allowing the method to produce surfaces with boundary as output.

\begin{figure}[htb]
  \centering
  \includegraphics[width=\linewidth]{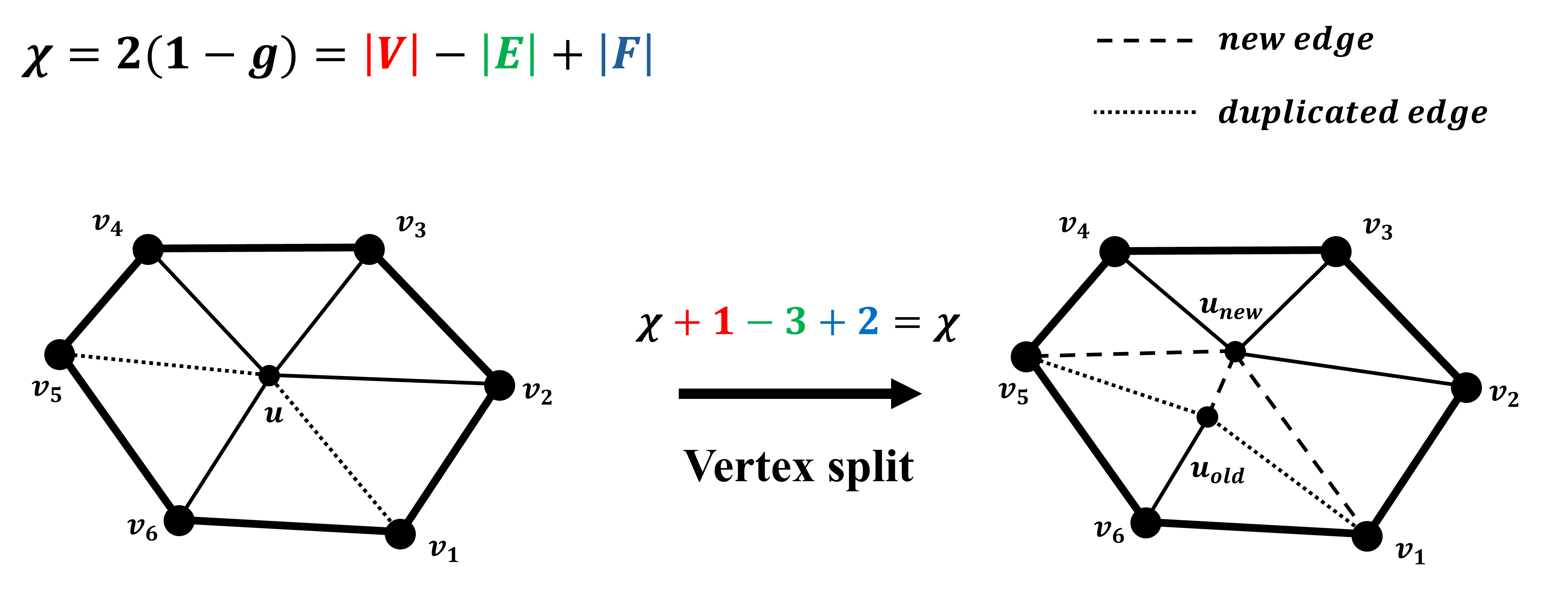}
  \caption{\label{fig:euler_op}
           Analysis of the Euler characteristic during the vertex split.}
\end{figure}

\subsection{Vertex split}
\label{sec:def}

We define one more Euler operator that restores the original point density after RsR, as shown in Figure \ref{fig:euler_op}.

Given a vertex $u$, the operator introduces a new edge $e=\{u_{old},u_{new}\}$ by splitting $u$ into two vertices and redistributing the incident connectivity accordingly. As shown in Figure  \ref{fig:euler_op}, two edges need to be selected as duplicated edges for the redistribution. Except for the edge $e$, two pairs of new edges and new faces need to be added. An example in Figure \ref{fig:euler_op} is candidate edge $\{u, v_4\}$. After the vertex split, a new edge \{$v_4, v_{new}$\} and a new face \{$v_4, u_{old}, u_{new}$\} are added. In total, we increase $|V|$ by 1, $|E|$ by 3, and $|F|$ by 2, resulting in an unchanged Euler characteristic.



\section{Method}
Figure \ref{fig:pipe} illustrates the pipeline of HRsR. Given a point cloud, HRsR first constructs a collapse graph, implemented as a $k$-nearest neighbor ($k$NN) graph. Edge collapses are then performed based on the connectivity defined by this graph. The corresponding collapse information is stored, and the resulting simplified point cloud is used as input to RsR for reconstruction. After reconstruction on this coarser point cloud, vertex split is applied to restore the mesh to its original vertex density according to the stored collapse information. For clarity, pre-processing steps on the input point cloud are omitted; they remain identical to those in RsR and are performed prior to the stages shown in the figure.

In the following, we will discuss how the two major novel components of HRsR, namely  \textit{edge collapse}, and \textit{vertex split} are employed by the algorithm.
\subsection{Edge collapse}
Given an input point cloud with normals, we first construct a collapse graph to simplify the data. This graph is defined as a $k$NN graph that provides local connectivity, on which subsequent edge collapse operations are performed. The process is controlled by two user-defined parameters: the number of neighbors $k$ and the collapse ratio $p$. Different from the $k$NN graph constructed in RsR, which ensures sufficient potential connections in the final reconstruction, HRsR uses a much smaller $k$ for collapsing, typically comparable to the size of a one-ring neighborhood, to improve efficiency. Given the collapse ratio $p$, we repeatedly apply edge collapses until only $p|V|$ vertices remain in the graph. Notably, to reduce the likelihood of edge crossings, edges are collapsed in ascending order of length, so that shorter edges are processed first. Additionally, edges connecting two vertices whose normal vectors form an angle greater than $90^\circ$ are not allowed to be collapsed. This constraint is essential to prevent collapsing vertices that belong to different surfaces. After each collapse operation, the previous coordinates and the new coordinate is stored sequentially and later used in the vertex split stage. The new position and normal of the merged vertex are computed as the weighted averages of the corresponding attributes of the two vertices. Edge collapses are performed subject to validity constraints and terminate automatically when no further valid edges remain.



\subsection{Vertex split}
In this phase, vertex splits are performed in the reverse order of the stored edge collapses, until either all collapses have been processed or the edge length falls below a user-specified threshold $l_{min}$. This strategy helps avoid most edge crossings while providing users with flexibility in controlling the output resolution, yielding meshes with the number of vertices ranging from $p|V|$ to $|V|$. However, this heuristic cannot completely eliminate edge crossings. Therefore, the intersection detection and resolution procedure described in Sec. \ref{sec:edge_crossing} is needed before every vertex split.

\begin{figure}[htb]
  \centering
  \includegraphics[width=\linewidth]{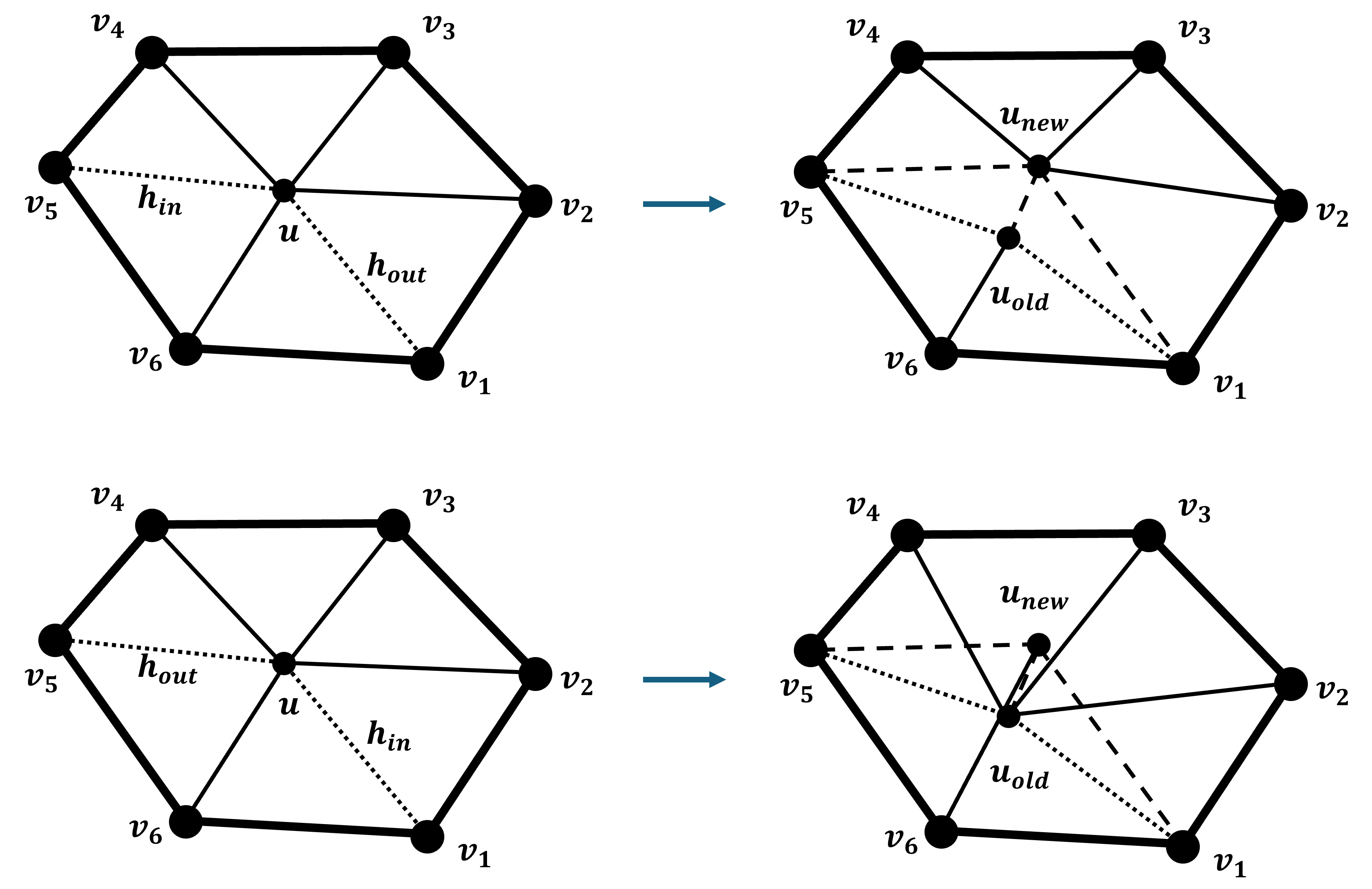}
  \caption{\label{fig:edge_selection}
            The ordering of edge candidates is meaningful when it comes to choosing the vertex split.}
\end{figure}

Furthermore, the selection of the two candidate 1-ring edges significantly affects the reconstruction quality. To determine a proper vertex split, we evaluate candidate pairs of incident halfedges $(h_{in}, h_{out})$ in the 1-ring neighborhood of a vertex by computing the dihedral angle and the minimum angle in the triangle. Figure \ref{fig:edge_selection} shows that swapping the assignments of  $h_{in}$ and $h_{out}$ results in different split outcomes. For each pair, we construct a local configuration induced by the candidate edge pair and assess its geometric quality via the following energy function,

\begin{equation}
    E = E_D+wE_A
\end{equation}

The dihedral term is defined as 

\begin{equation}
    E_D  = \sum_{(t_i, t_j)} \left( \langle \mathbf{n}_i, \mathbf{n}_j \rangle - 1 \right)^2 \, \ell_{ij}
\end{equation}

where $\mathbf{n}_i, \mathbf{n}_j$ are triangle normals and $\ell_{ij}$ is the shared edge length. This formulation favors aligned normals and penalizes sharp folds. 

The minimum angle term is defined as

\begin{equation}
E_{A} = \sum_{t} \, \ell_{\min}(t) \, \left( \frac{\pi}{3} - \theta_{\min}(t) \right)
\end{equation}

where $\theta_{\min}(t)$ is the smallest angle of triangle $t$, $\ell_{min}(t)$ is the shortest edge length. 


The optimal edge pair is selected by minimizing $E$ over all ordered edge pairs in the vertex's one-ring. To avoid degenerate configurations, we additionally enforce a constraint on the maximum dihedral angle and fall back to the best feasible alternative when necessary.

\begin{figure}[htb]
  \centering
  \begin{subfigure}{0.48\linewidth}
    \centering
    \includegraphics[width=\linewidth]{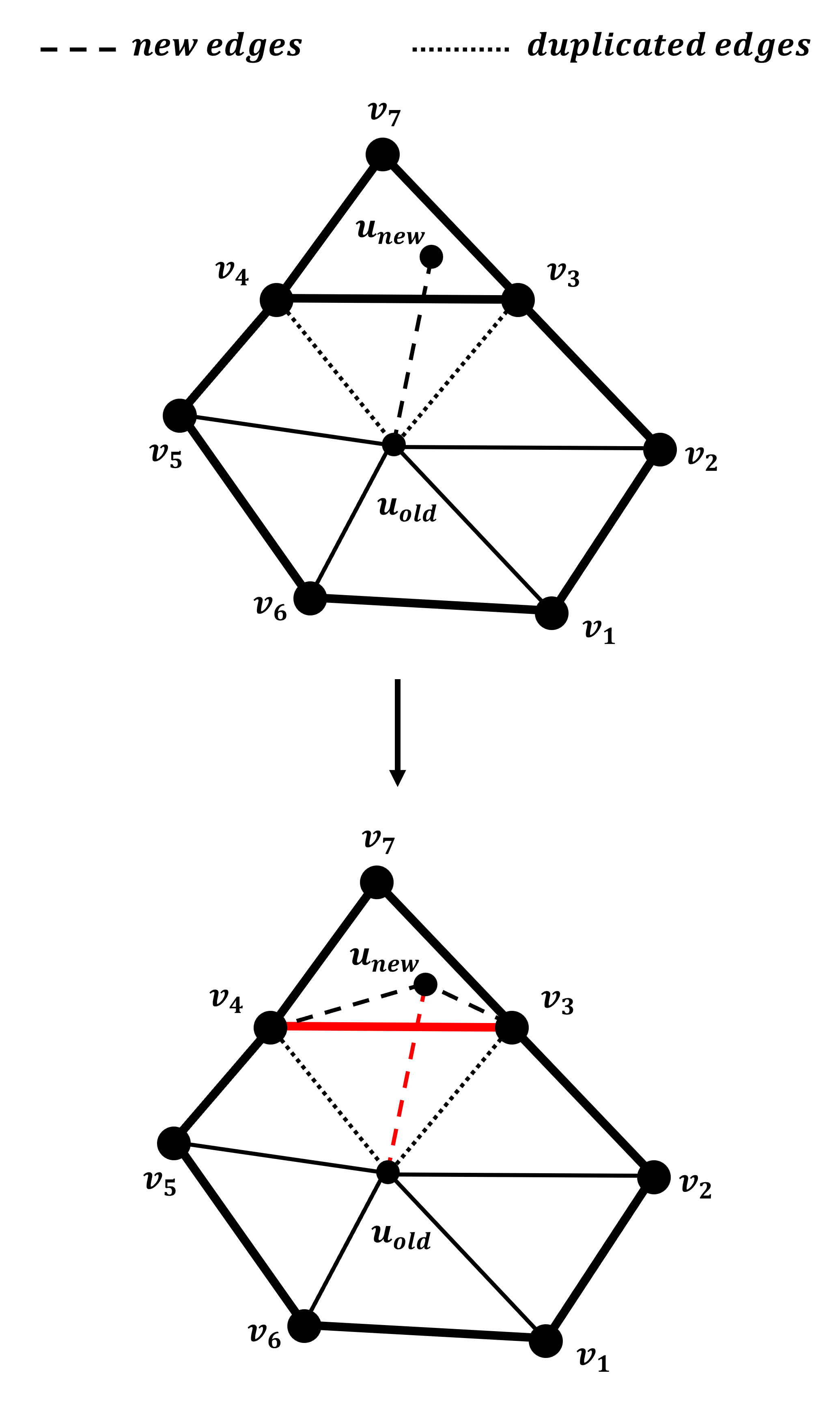}
    \caption{The crossing edges in red will cause triangle intersections after the vertex split.}
    \label{fig:edge_crossing}
  \end{subfigure}
  \hfill
  \begin{subfigure}{0.48\linewidth}
    \centering
    \includegraphics[width=\linewidth]{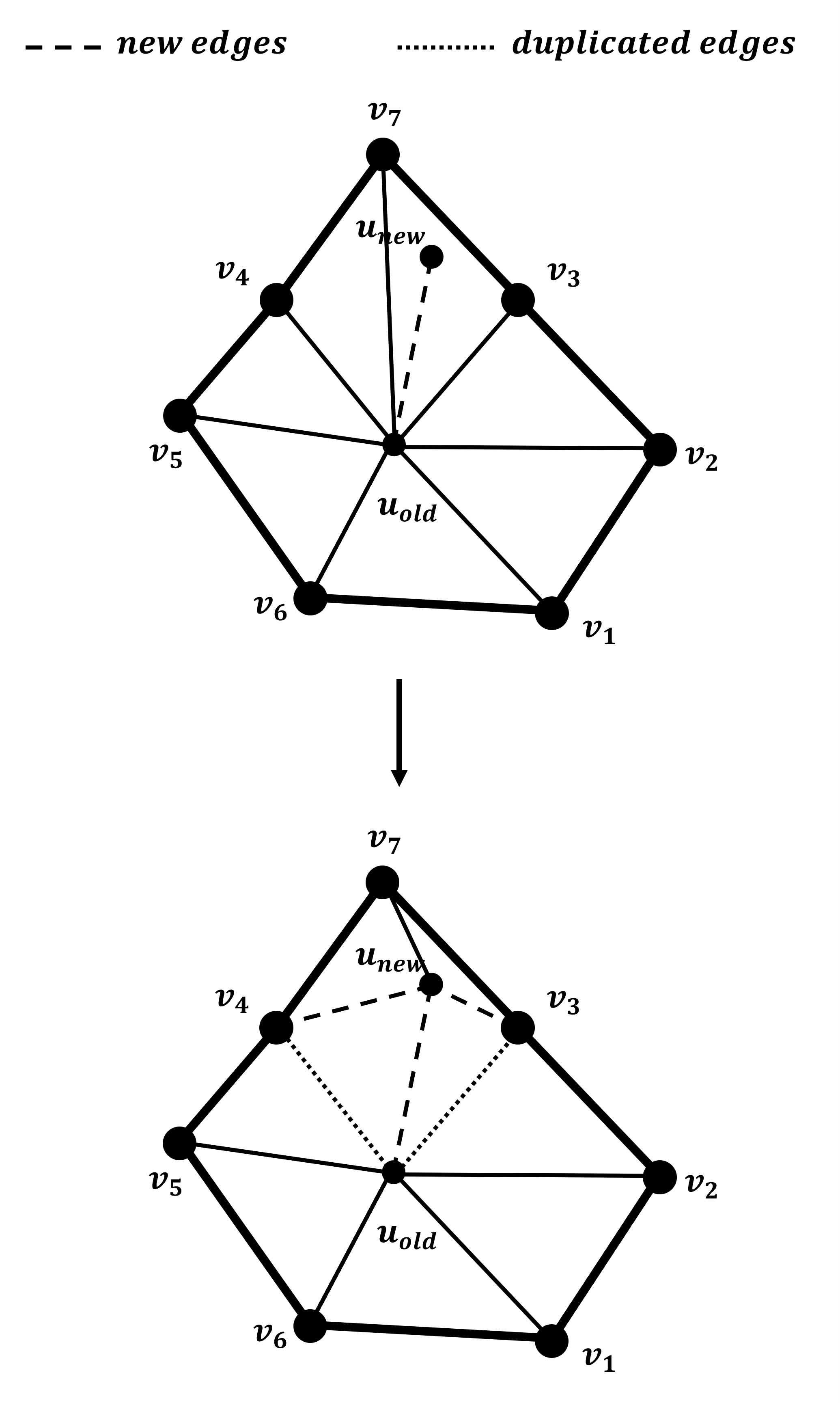}
    \caption{By flipping the intersecting edge, the potential intersection is resolved.}
    \label{fig:edge_crossing_2}
  \end{subfigure}

  \caption{Resolution of edge crossing. Note both $u_{old}$ and $u_{new}$ might cause intersections.}
  \label{fig:edge_crossing_combined}
\end{figure}

\begin{figure}[htb]
  \centering
  \includegraphics[width=\linewidth]{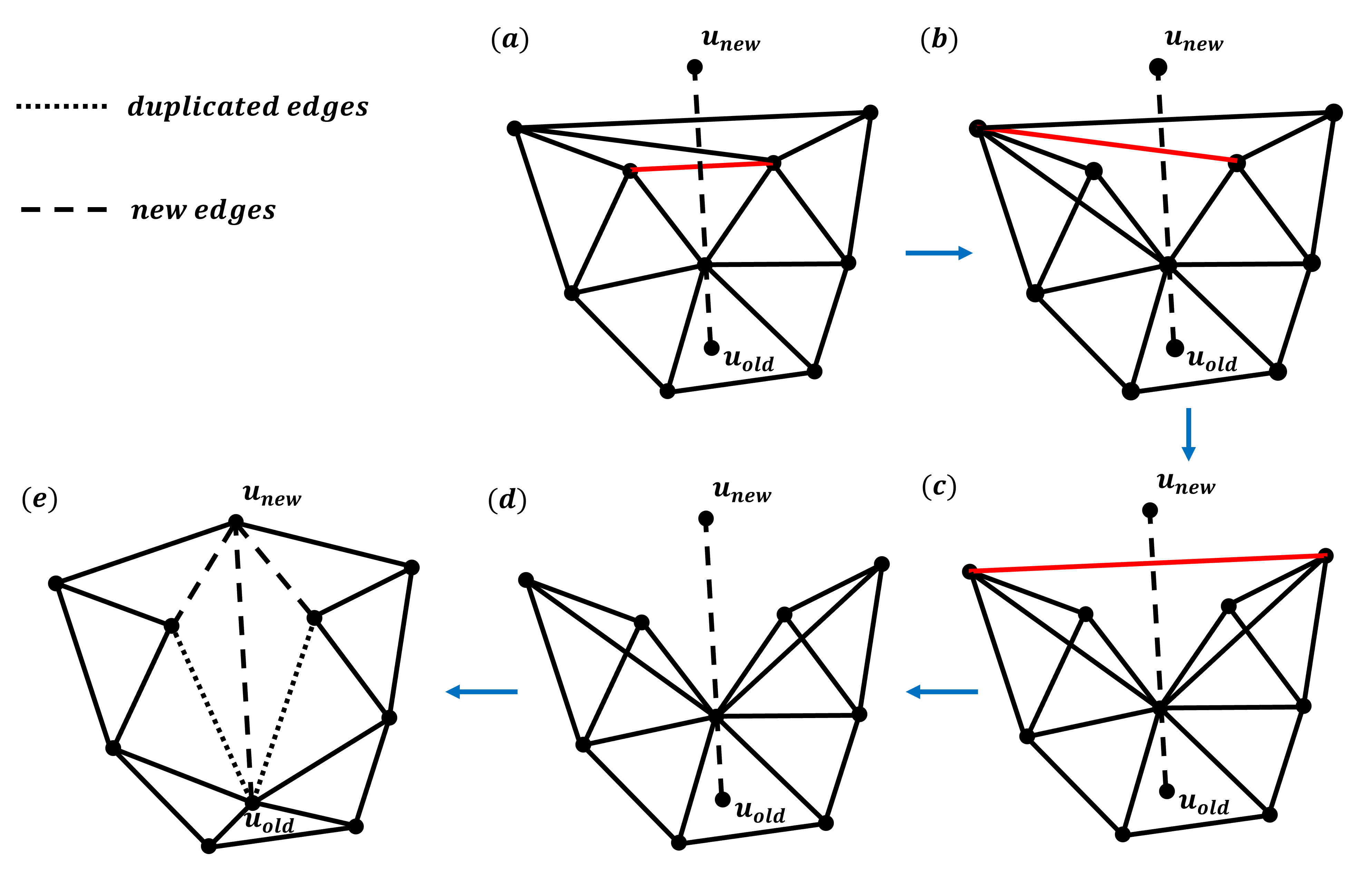}
  \caption{\label{fig:edge_crossing_bdry}
           Resolution for corner cases where edge flipping cannot work. In this figure, we also show a more complicated case with multiple intersections. The detected crossing edge in each stage is marked red. }
\end{figure}
\subsubsection{Edge crossing}
\label{sec:edge_crossing}

Although vertex splits preserve topology, they may introduce geometric inconsistencies. Figure \ref{fig:edge_crossing} illustrates such a problematic configuration in 2D.

Specifically, if the new edge \{$u_{new}, u_{old}$\} created by the vertex split does not lie entirely within the 1-ring neighborhood, face intersections may occur. To address this issue, we first detect intersections between the new edge and the boundary edges of the 1-ring triangle fan (e.g., edge \{$v_3, v_4$\}). If an intersection is detected, we perform an edge flip on the corresponding boundary edge. This operation effectively enlarges the 1-ring neighborhood so that the new edge is fully contained within it, as illustrated in Figure \ref{fig:edge_crossing_2}. 

A corner case arises when the newly created edge intersects a boundary edge that cannot be resolved, since boundary edges cannot be flipped, as illustrated in Figure \ref{fig:edge_crossing_bdry}.$(a)$. In this situation, we first perform edge flips to resolve all interior edge intersections until the intersection with the boundary edge is encountered. We then remove the intersecting boundary edge to eliminate the conflict. 

In 3D, not all edge intersections in the tangent plane necessarily correspond to face intersections in the embedded surface. However, following the convention of RsR, we adopt a stricter notion of geometric consistency by projecting all primitives onto the tangent plane defined by the vertex normal and resolving intersections in this domain.

It is worth noting that this design also contributes to the observed speedup. Compared to the geometry test in RsR, our consistency handling is more lightweight, as it is applied after reconstruction on a simplified point cloud. At this stage, we operate on a well-defined mesh with a half-edge representation, rather than iterating over all possible candidate edges in a spherical search space.


\begin{figure*}[htb]
  \centering
  \includegraphics[width=\linewidth]{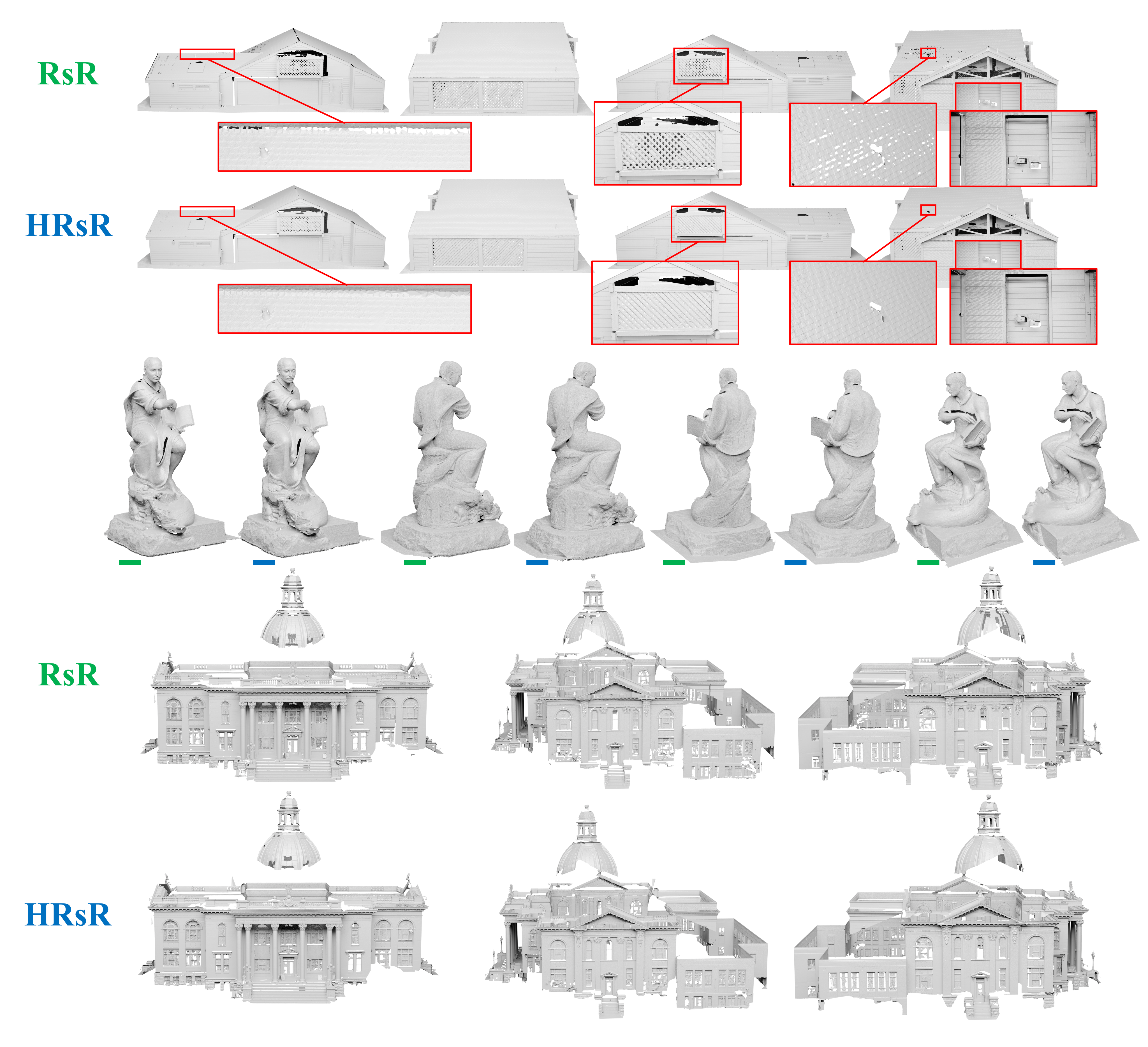}
  \caption{\label{fig:tanks}
           Qualitative comparison between RsR and HRsR on \textit{Tanks and Temple}. Meshes reconstructed by RsR are indicated in green, while those reconstructed by HRsR are indicated in blue.}
\end{figure*}

\begin{figure*}[htb]
  \centering
  \includegraphics[width=\linewidth]{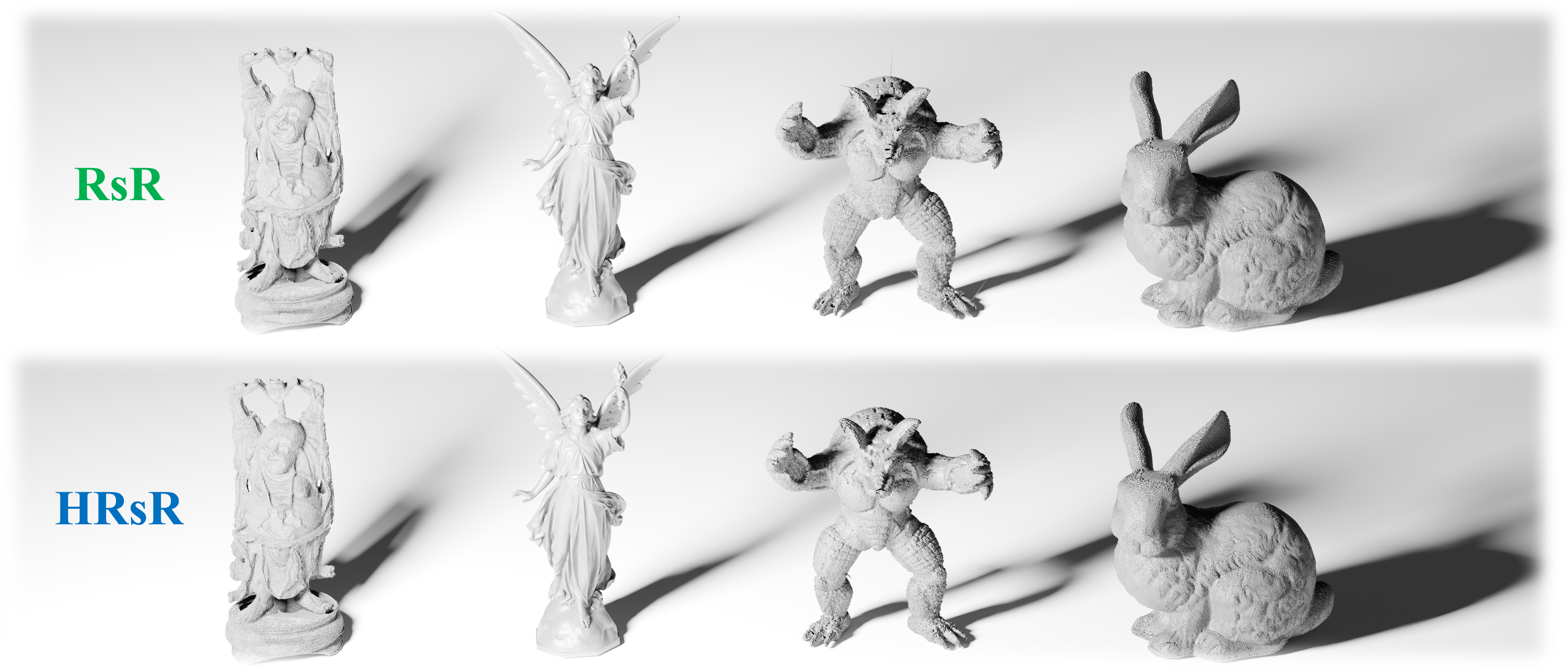}
  \caption{\label{fig:stanford}
           Qualitative comparison between RsR and HRsR on \textit{Stanford 3D Scanning Repository}.}
\end{figure*}
%
\subsubsection{Local optimization}
%
After each vertex split, we perform an iterative connectivity optimization through a series of edge flips. Candidate edges are collected from the neighborhood of the newly inserted vertex by traversing its 1-ring, the outer loop edges around the vertex, as well as the edges of the triangles of the outer loop.


In each refinement iteration, the candidate edges are sorted in ascending order of edge length and processed in reverse order, prioritizing longer edges that are more likely to produce poorly shaped triangles. For each edge, we evaluate whether an edge flip is beneficial using a local angle-based criterion. An edge flip is performed if all four of the angles within the two triangles sharing that edge are below 90 degrees and any of them are below a user-supplied parameter. The first condition prevents flips from creating degenerate geometry, while the second condition provides a fast and effective heuristic for worthwhile edge flips.



\section{Implementation}
Our implementation is based on the official RsR implemented in GEL, a geometry processing library with a self-contained half-edge representation. Our code is available at \url{https://github.com/TO_BE_OPEN_SOURCED}.

\section{Results}
All experiments are conducted on large-scale real-world scanning datasets. The same RsR parameters are applied in different experiments using tangent distance: $k=70, r=20, \theta=60, n=50$, where $k$ denotes the number of neighbors, $r$ is the threshold for removing far away vertices, $\theta$ denotes the threshold of normal difference for rejecting neighbor connection, and $n$ defines the threshold for distinguishing spurious handles from real ones. We refer readers to the original paper for the formal definition of the parameters. In particular, we use the \textit{Stanford 3D Scanning Repository} and \textit{Tanks and Temples} \cite{Knapitsch2017} as representative benchmarks, providing detailed quantitative evaluations of efficiency as well as qualitative comparisons with the original RsR. We use the \textit{DTU Robot Image Dataset} \cite{jensen2014large} to compare with Co3Ne \cite{boltcheva2017surface} and demonstrate that HRsR preserves the desirable properties of RsR. All experiments were conducted on an HPC system using Intel Xeon E5-2620 v4 processors, with 20 CPU cores allocated per run and 256 GB of RAM.

\subsection{Representative datasets}

The number of points in the representative datasets ranges from 0.36M to 68M. However, at the original resolution, RsR frequently triggers memory swapping, leading to significant performance degradation. To ensure we are not merely benchmarking the swap performance of our system on larger models, we downsample datasets containing more than 20M points to 20M points. Table \ref{tab:speed} and Table \ref{tab:space} present quantitative comparisons in terms of both runtime and memory usage. Figure \ref{fig:tanks} and Figure \ref{fig:stanford} illustrate the qualitative comparison between RsR and HRsR. We use the default number of neighbors, which is 6, for constructing the $k$NN graph in the edge collapse of HRsR.

As seen in\ref{tab:speed}, HRsR achieves up to $3\times$, $6\times$, and $6.5\times$ speedups at 50\%, 25\%, and 12.5\% collapse levels, respectively. We also include a comparison with an improved version of RsR that introduces parallelism where possible, e.g., in the normal estimation procedure. We report these measurements to ensure transparency in the evaluation. All speedups are computed using the original RsR implementation in GEL as the baseline. Table \ref{tab:space} shows that HRsR achieves a significant reduction in memory usage compared to RsR. At the 12.5\% collapse level, HRsR reduces memory consumption by approximately $7.5\times$, enabling the reconstruction of larger meshes on memory-limited hardware.

Qualitatively, Figures \ref{fig:tanks} and \ref{fig:stanford} demonstrate that HRsR produces reconstruction results comparable to those of RsR. Zoomed-in views of the \textit{Barn} mesh are provided for further inspection. The visualizations of HRsR on the \textit{Tanks and Temple} scenes are rendered at a collapse level of 12.5\%, which represents the most challenging setting for vertex split. The same setting applies to \textit{Stanford 3D Scanning Repository} as well.


Meanwhile, these comparisons reveal an additional characteristic of HRsR that resembles a hole-filling effect. This behavior arises because edge collapse effectively enlarges the spatial neighborhood of the $k$NN graph, given the same parameter $k$ used in both RsR and HRsR. However, this effect is not always desirable. For instance, fine structures such as openwork railings in the Barn model can be closed. Therefore, users should take this trade-off into consideration when selecting reconstruction methods.



\begin{table*}[t]
\centering
\begin{tabularx}{\linewidth}{ c | Y | Y | Y | Y | Y | Y | Y | Y | Y}
\hline
\multirow{2}{*}{Meshes} & \multicolumn{3}{c|}{Co3Ne w/o smoothing} & \multicolumn{3}{c|}{Co3Ne w/ smoothing} & \multicolumn{3}{c}{HRsR} \\
\cline{2-10}
 & $r_v(\%)$ & $|E_b|$ &  $t(s)$ & $r_v(\%)$ & $|E_b|$ &  $t(s)$ & $r_v(\%)$ & $|E_b|$ &  $t(s)$ \\
\hline
stl002     &87.067         &1.65M      &94      & 99.994      & 2,902      &\textbf{21}   &\textbf{99.999}  & \textbf{1,218}  &930\\
\hline
stl003     &86.598         &1.59M      &92     & 99.741      &29,538       &\textbf{20}   &\textbf{99.974}   &\textbf{4,097}   &939\\
\hline
stl024     &89.306      &2.56M      &177  & 99.919      &19,287       &\textbf{37}    &\textbf{99.996}   &\textbf{4,484}  &1,670\\
\hline
\end{tabularx}
\caption{\label{tab:dtu}Quantitative comparisons between Co3Ne and HRsR on meshes from \textit{DTU Robot Image Dataset}. The best performances are in boldface.}
\end{table*}

\subsection{DTU Robot Image Dataset}
\begin{figure}[htb]
  \centering
  \includegraphics[width=\linewidth]{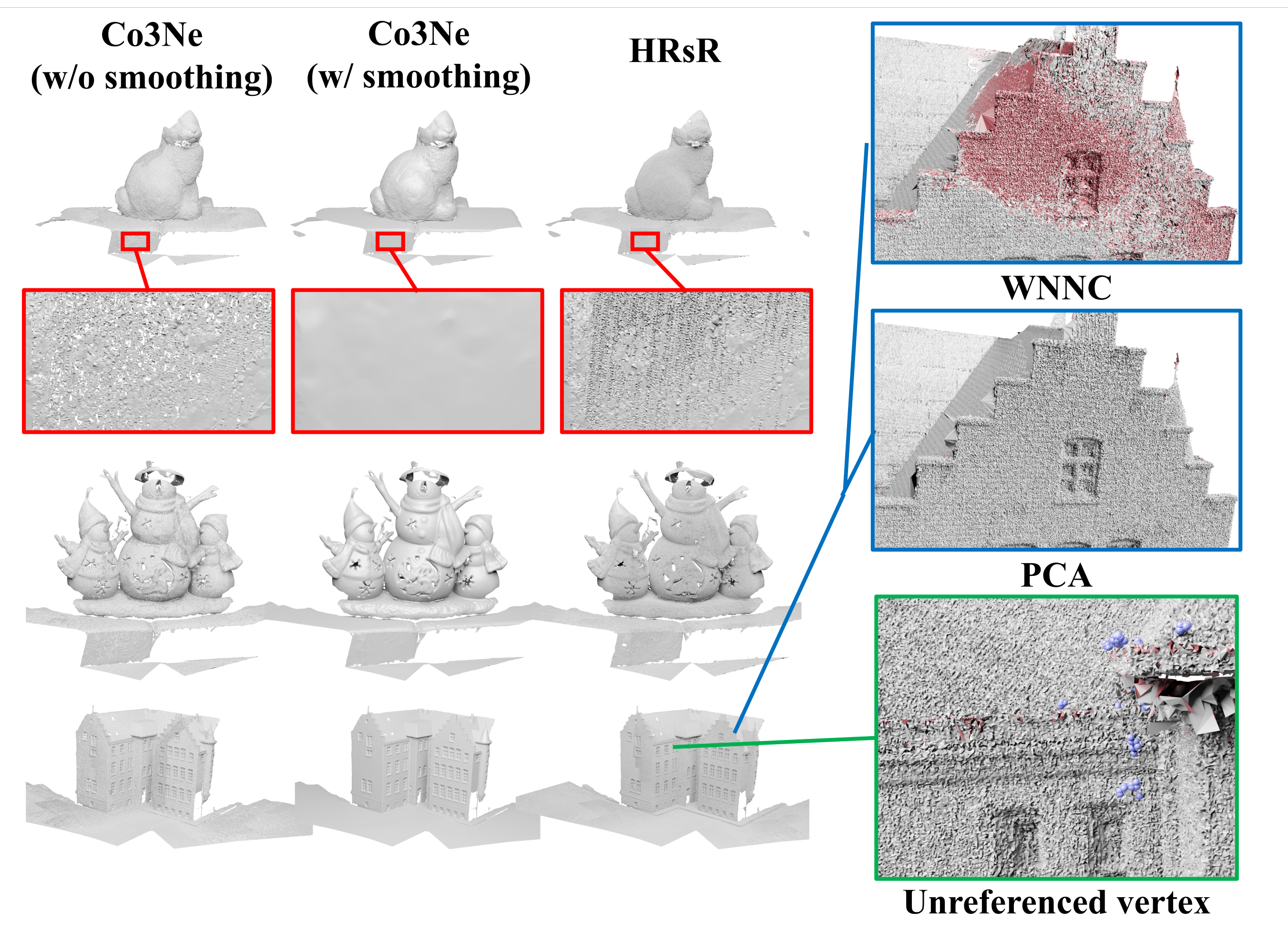}
  \caption{\label{fig:dtu}
           Qualitative results on \textit{stl002, stl003}, and \textit{stl024}. On the right side, we show a comparison using plain normal estimation and the advanced normal estimation method, highlighted with a blue border. The backfaces are rendered in red. The bottom-right figure, highlighted with a green border, shows vertices (rendered as blue spheres) that are not included in HRsR.}
\end{figure}

To demonstrate that HRsR preserves the desirable properties of RsR, we conduct experiments on meshes from the \textit{DTU Robot Image Dataset} \cite{jensen2014large}. We compare against Co3Ne \cite{boltcheva2017surface}, a state-of-the-art combinatorial reconstruction method prior to RsR. Figure \ref{fig:dtu} presents qualitative comparisons on meshes \textit{stl002}, \textit{stl003}, and \textit{stl024}. Two versions of Co3Ne reconstruction are presented. Co3Ne w/ smoothing reconstructs meshes from a smoothed point cloud, yielding visually pleasing results but not preserving the original vertex positions. Co3Ne w/o smoothing failed on these scanning data. A zoomed-in view of the bottom region of the animal mesh is provided to illustrate representative reconstruction quality.

We adopt the same quantitative metrics as in RsR, namely the vertex reference ratio $r_v$ and the number of boundary edges $|E_b|$. Specifically, $r_v$ is calculated by $\frac{|V_o|}{|V_i|}$, where $|V_i|$ denotes the number of points as input, and $|V_o|$ denotes the number of \textit{original} points referenced in the output mesh. We assume vertices in Co3Ne w/ smoothing can also be projected back to the original position and evaluate comparisons with both versions of reconstruction by Co3Ne. Table \ref{tab:dtu} shows the evaluation results, which support the claim that HRsR preserves the desirable properties of RsR.

In addition, we further test our method with a recent advanced normal estimation method, WNNC \cite{lin2024fast}, on this data. Figures on the right side of Figure \ref{fig:dtu} with blue boundaries illustrate the comparison between WNNC and plain PCA \cite{hoppe1992surface} used in our method. The backfaces are rendered in red. Default parameters with a noise level $l5$ are used for WNNC. It turns out that WNNC does not work well on this data.

On the bottom right of Figure \ref{fig:dtu}, we visualize examples of vertices that are not included in the reconstruction from HRsR as blue spheres. These points are either small clusters of points whose size is below the component size threshold to be reconstructed, or single floating points whose normal is inconsistent with all their neighbors.

\subsection{Filtering}

\begin{figure}[htb]
  \centering
  \includegraphics[width=\linewidth]{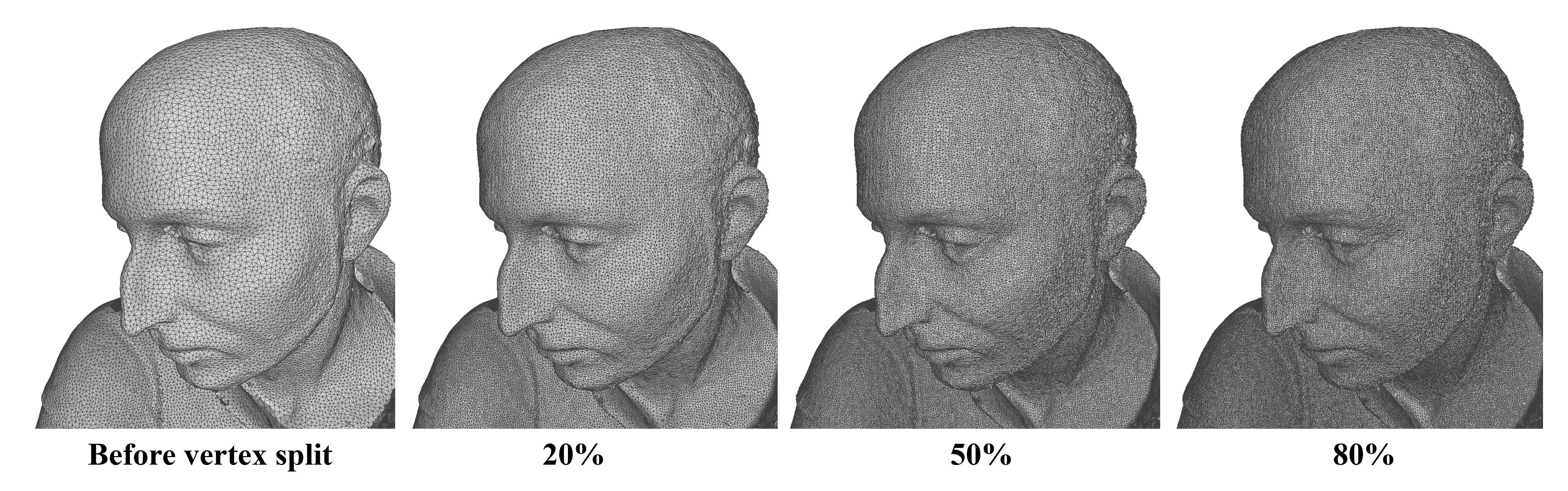}
  \caption{\label{fig:progressive}
           Output meshes at different stages of vertex split. }
\end{figure}

Another advantage of HRsR is that it provides users with the flexibility to control the reconstruction under different levels of point density. In some scenarios, strict information preservation is not a primary concern. In such cases, a user-defined parameter can be used to determine when vertex splitting is terminated and reconstruction is completed. Figure \ref{fig:progressive} illustrates reconstructed meshes under different point density levels.

\begin{figure}[htb]
  \centering
  \includegraphics[width=\linewidth]{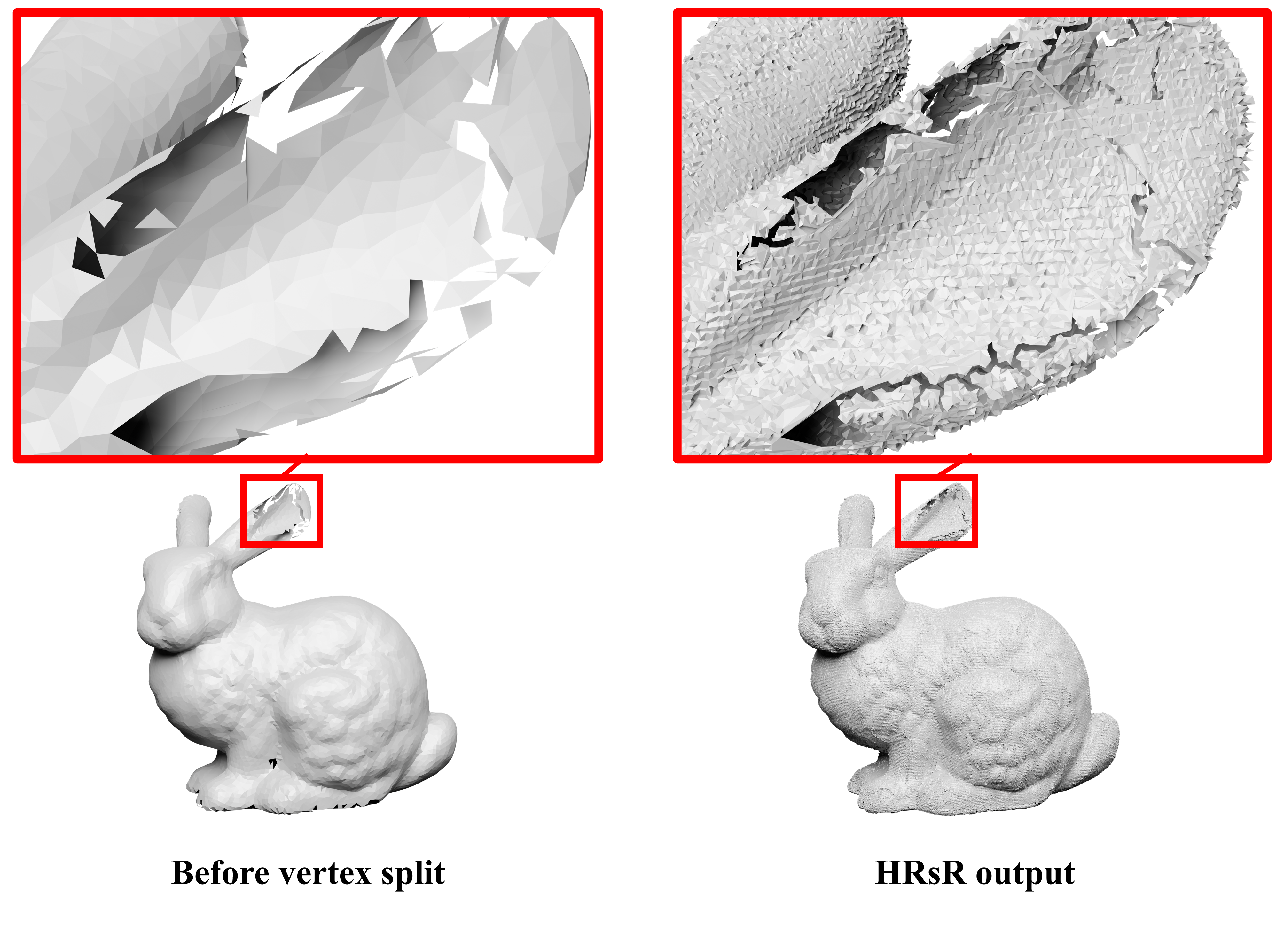}
  \caption{\label{fig:fail}
           The failed region when the collapse ratio is set to 2.7\%. }
\end{figure}
\subsection{Failure case}
It is reasonable to ask whether the benefits of edge collapse have been fully exploited in our experiments. 
Unfortunately, when the point cloud becomes too sparse, RsR fails to reconstruct high-curvature regions. Figure \ref{fig:fail} shows an example of the \textit{Stanford Bunny} at a collapse ratio of 2.7\%. This is simply because the normals are too different at that point. Moreover, we have observed that we get diminishing returns in terms of performance with more than five iterations of edge collapses. 

\begin{table*}[t]
\centering
\begin{tabularx}{\linewidth}{ c | Y | Y | Y Y Y}
\hline
\multirow{2}{*}{Meshes ($|V|$)} & \multirow{2}{*}{RsR} & \multirow{2}{*}{\begin{tabular}{c}RsR\\(parallel)\end{tabular}} & \multicolumn{3}{c}{HRsR} \\
\cline{4-6}
 & & & 50\% & 25\% & 12.5\% \\
\hline
Barn (12.75M)           & 9493 (2.64h)      & 6973 (1.93h)      & 3622 (2.62x)  & 2601 (3.65x)  & 2032 (4.67x) \\
Caterpillar (7.26M)     & 3855 (1.07h)      & 2684 (0.75h)      & 1824 (2.11x)  & 1323 (2.91x)  & 1085 (3.55x) \\
Church (20M)                        & 21987 (6.11h)     & 14838 (4.12h)     & 7344 (2.99x)  & 3583 (6.13) & 3400 (6.47x)\\
Courthouse (20M)                    & 21019 (5.84h)     & 14623 (4.06h)     & 6838 (3.07x)  & 4291 (4.90x)  & 3355 (6.26x)\\
Ignatius (5.02M)        & 3433 (0.95h)      & 2127 (0.59h)      & 1318 (2.60x)  & 924 (3.72x)  & 726 (4.73x)\\
Meetingroom (20M)                   & 20429 (5.67h)     & 17211 (4.78h)     & 7339 (2.78x)   & 4470 (4.57x)  & 3389 (6.03x)\\
Truck (8.07M)           & 4547 (1.26h)      & 3369 (0.93h)      & 2158 (2.11x)  & 1540 (2.95x)  & 1237 (3.68x)\\
\hline
\hline
Buddha (2.56M)          & 1894 (0.53h)     & 1220 (0.34h)      & 812 (2.33x)  & 653 (2.90x)  & 576 (3.29x) \\

Armadillo (0.61M)       & 388      & 238      & 158 (2.46x)  & 129 (3.01x)  & 121 (3.21x) \\

Bunny (0.36M)           & 208      & 116      & 81 (2.57x)  & 68 (3.06x)  & 61 (3.41x) \\

Lucy (14.03M)           & 9202 (2.56h)      & 5158 (1.43h)      & 3891 (2.36x)  & 3345 (2.75x)  & 3033 (3.03x) \\

Statue (5.00M)          & 3022 (0.84h)      & 1788 (0.50h)      & 1380 (2.19x)  & 1159 (2.61x)  & 1052 (2.87x) \\
\hline
\end{tabularx}
\caption{\label{tab:speed}Execution time of RsR, RsR (parallel), and HRsR on different meshes (in seconds). RsR (parallel) denotes an optimized implementation of RsR without algorithmic modifications. The top half of the table reports results on the \textit{Tanks and Temple} dataset, while the bottom half shows results on the \textit{Stanford 3D Scanning Repository}. For time evaluations of RsR and RsR (parallel), we report the equivalent runtime in hours in parentheses for better interpretability. In contrast, for HRsR, we report the speedup relative to RsR in parentheses.}
\end{table*}

\begin{table*}[t]
\centering
\begin{tabularx}{\linewidth}{ c | Y | Y | Y Y Y}
\hline
\multirow{2}{*}{Meshes ($|V|$)} & \multirow{2}{*}{RsR} & \multirow{2}{*}{\begin{tabular}{c}RsR\\(parallel)\end{tabular}} & \multicolumn{3}{c}{HRsR} \\
\cline{4-6}
 & & & 50\% & 25\% & 12.5\% \\
\hline
Barn (12.75M)           & 100.43     & 72.45      & 38.86  & 20.54  & 13.45 \\
Caterpillar (7.26M)     & 32.90      & 27.19      & 14.17  & 8.84  & 7.40 \\
Church (20M)                        & 155.48     & 104.20     & 54.15  & 29.13  & 18.30 \\
Courthouse (20M)                    & 152.96     & 91.95      & 48.15  & 24.53  & 18.30 \\
Ignatius (5.02M)        & 38.76      & 29.36      & 15.70  & 7.92  & 4.66 \\
Meetingroom (20M)                   & 154.70     & 116.64     & 62.28  & 31.76  & 18.44\\
Truck (8.07M)           & 41.78      & 30.51      & 17.08  & 10.51  & 7.93 \\
\hline
\hline
Buddha (2.56M)            & 19.93      & 16.31      & 11.69  & 4.25  & 3.16 \\

Armadillo (0.61M)            & 4.85      & 3.90      & 2.35  & 1.12  & 0.86 \\

Bunny (0.36M)          & 2.93      & 2.20      & 1.61  & 0.72  & 0.43 \\

Lucy (14.03M)         & 108.44      & 81.44      & 36.59  & 21.74  & 15.94 \\

Statue (5.00M)           & 38.61      & 31.91      & 18.37  & 8.40  & 4.79 \\
\hline
\end{tabularx}
\caption{\label{tab:space}Peak runtime memory consumption of RsR, RsR (parallel), HRsR on different meshes (in GB). The top half of the table reports results on the \textit{Tanks and Temple} dataset, while the bottom half shows results on the \textit{Stanford 3D Scanning Repository}.}
\end{table*}


\section{Discussion}
We have presented HRsR, a version of the RsR method which is both much faster and much more space efficient while preserving the key properties of the original algorithm. A practical benefit of these efficiency gains is that we can apply HRsR to much larger data on the same hardware.

HRsR is based on a hierarchical scheme which, in principle, could be applied together with any other combinatorial reconstruction method. The choice of RsR is motivated in part by the fact that this algorithm incorporates almost all points in its reconstruction. Points are only omitted in rare cases where the rotation system is not consistent with connected points. This makes RsR and HRsR useful in cases where it is important to avoid the bias that could be introduced by arbitrary omission of input points. 
Another motivating factor is that we inherit the topology control properties of RsR. In particular, RsR and HRsR do not introduce spurious handles since handles are always added explicitly. This can be helpful in eliminating topological noise. 
%
%
\subsection{Future Work}
As the hierarchy of edge collapses and splits rely purely on local information with no global invariants, HRsR is friendly to spatial partitioning schemes. We plan to explore this in the future. Particular challenges would include efficient spatial partitioning and boundary handling. Modifications of the output structure is also required.


As another future direction, the mechanism for topology control (user specified number of handles) offered by RsR/HRsR is quite different from the persistence-based schemes which rely on filtrations. It would be interesting to compare the two approaches. 

\bibliographystyle{eg-alpha-doi} 
\bibliography{egbibsample}       




\end{document}